\documentclass[10pt,conference]{IEEEtran}
\usepackage[centertags]{amsmath}
\usepackage{amsfonts}
\usepackage{amssymb}

\newcommand{\X}{{\mathbf X}}
\newcommand{\M}{{\mathbf M}}
\newcommand{\Y}{{\mathbf Y}}
\newcommand{\prob}{\text{Pr}}

\newcommand{\peab}{P_{e|{\cal A},{\cal B}}} 
\newcommand{\qed}{\hspace*{\fill}~{\rule{2mm}{2mm}}\par\endtrivlist\unskip}
\newcommand{\tends}{\rightarrow}

\newcommand{\prcavg}{\bar{P}_{e,rc}}
\newcommand{\pe}{\bar{P}_{e}}

\newtheorem{theorem}{Theorem}
\newtheorem{corollary}{Corollary}

\begin{document}

\title{A Channel Coding Perspective of Recommendation Systems}
\author{
\authorblockN{S.T. Aditya}
\authorblockA{Department of Electrical Engineering\\
Indian Institute of Technology Bombay \\
Mumbai, India\\
Email: staditya@ee.iitb.ac.in}
\and
\authorblockN{Onkar Dabeer}
\authorblockA{School of Technology \\
and Computer Science \\
Tata Institute of Fundamental Research\\
Mumbai, India \\
Email: onkar@tcs.tifr.res.in}
\and
\authorblockN{Bikash Kumar Dey}
\authorblockA{Department of Electrical Engineering \\
Indian Institute of Technology Bombay \\
Mumbai, India\\
Email: bikash@ee.iitb.ac.in}
}
\maketitle

\begin{abstract}
Motivated by recommendation systems, we consider the problem of estimating block constant binary matrices (of size $m \times n$) from sparse and noisy observations. The observations are obtained from the underlying block constant matrix after unknown row and column permutations, erasures, and errors. We derive upper and lower bounds on the achievable probability of error. For fixed erasure and error probability, we show that there exists a constant $C_1$ such that if the cluster sizes are less than $C_1 \ln(mn)$, then for any algorithm the probability of error approaches one as $m, n \tends \infty$. On the other hand, we show that a simple polynomial time algorithm gives probability of error diminishing to zero provided the cluster sizes are greater than $C_2 \ln(mn)$ for a suitable constant $C_2$.       
\end{abstract}

\section{Introduction}

Recommender systems are commonly used to suggest content (movies, books, etc.) that is relevant to a given buyer. The most common approach is to {\it predict} the rating that a potential buyer might assign to an item and use the predicted ratings to recommend items. The problem thus reduces to completion of the rating matrix based on a sparse set of observations. This problem has been popularized by the Netflix Prize (\cite{netflix}). A number of methods have been suggested to solve this problem; see for example \cite{surveypaper}, \cite{sp.issue}, \cite{koren} and references therein. 
Recently, several authors (\cite{recht1}, \cite{recht2} \cite{montanari}) have used the assumption of a low-rank rating matrix to propose provably good algorithms.  For example, in \cite{recht1}, \cite{recht2}, a ``compressed sensing" approach based on nuclear-norm minimization is proposed. It is shown in \cite{recht2} that if the number of samples is larger than a lower bound (depending on the matrix size and rank), then with high probability, the proposed optimization problem exactly recovers the underlying low-rank matrix from the samples. In \cite{montanari}, the relationship between the ``fit-error" and the prediction error is studied for large random matrices with bounded rank. An efficient algorithm for matrix completion is also proposed. 

In this paper, we consider a different setup. We assume that there is an underlying ``true" rating matrix, which has block constant structure. In other words, buyers (respectively items) are clustered into groups of similar buyers (respectively items), and similar buyers rate similar items by the same value. The observations are obtained from this underlying matrix (say $\M$) as described below.
\begin{enumerate}
\item The rows and columns of $\M$ are permuted with unknown permutations, that is, the clusters are not known.
\item Many entries of $\M$ are erased by a memoryless erasure channel. This models the sparsity of the available ratings.
\item The non-erased entries are observed through a discrete memoryless channel (DMC). This channel models 
\begin{itemize}
\item the residual error in the block constant model, and,
\item the ``noisy" behavior of buyers who may rate the same item differently at different times.
\end{itemize}
\end{enumerate}
One may also treat these two channels as a single effective DMC, but we prefer the above break-up for
conceptual reasons.
Our goal is to identify conditions on the cluster sizes under which the underlying matrix can be recovered with small probability of error.  Our recommendation system model differs from \cite{recht1}, \cite{recht2}, and in particular, we do not seek completion of the observed matrix, but rather the recovery of the underlying $\M$. As described above, our goal reduces to analyzing the error performance of the code of block-constant matrices over the channel described above. 

From a practical stand-point, it is desirable to consider the case when the parameters of the erasure channel and DMC are not known. However, in this paper, we consider the simpler case when these channel parameters are known. In particular, for simplicity, we consider the case when $\M$ is an $m\times n$ matrix with entries in $\{0,1\}$ and the DMC is a binary symmetric channel (BSC) with error probability $p$. The erasure probability is $\epsilon$. Our main results are of the following nature.
\begin{itemize}
\item If the ``largest cluster size" (defined precisely in Section \ref{sect:main}) is less than $C_1 \ln(mn))$, then the probability of error approaches unity for {\it any} estimator of $\M$ as $mn \tends \infty$ (Corollary \ref{cor:cor2}, Part 2)).
\item We analyze a simple algorithm, which clusters rows and columns first, and then estimates the cluster values. We show that if the ``smallest cluster size" is greater than a constant multiple of $\ln(mn)$, then the probability of error for this algorithm (averaged over the rating matrices), approaches zero as $mn \tends \infty$ (Theorem \ref{thm:final}, Part 2)). Combined with the previous result, this implies that $\ln(mn)$ is a sharp threshold for exact recovery asymptotically. 
\item If we consider the probability of error for a fixed rating matrix, then the algorithm needs the smallest cluster size to be larger than a constant multiple of $\sqrt{mn\ln(m) \ln(n)}$.
\end{itemize}
While we obtain the asymptotic results for fixed $p$ and $\epsilon$, the bounds we obtain in the process also apply to the case when $p$, $\epsilon$ depend on $m$, $n$.

The paper is organized as follows. In Section \ref{sect:model}, we describe our model. The main results are stated and proved in Section \ref{sect:main}. We conclude in Section \ref{sect:con}.

\section{Our Model and Notation}
\label{sect:model}

Suppose $\X$ is the unknown $m \times n$ rating matrix with entries in $\{0,1\}$, where $n$ is the number of buyers and $m$ is the number of items. Let ${\cal A}=\{A_i\}_{i=1}^r$ and ${\cal B}=\{B_j\}_{j=1}^t$ be partitions of $[1:m]$ and $[1:n]$ respectively. The sets $A_i \times B_j$ are the clusters in the matrix $\X$. We call $A_i$'s ($B_j$'s) the row (column) clusters. We denote the corresponding row and column cluster sizes by $m_i$ and $n_j$, and the number of row clusters and the number of column clusters by $r$ and $t$ respectively. (We note that the $A_i$'s (respectively $B_i$'s) need not consist of adjacent rows (respectively columns) and hence this notation is different
from that in the introduction). The entries of $\X$ are passed through the cascade of a memoryless erasure channel with erasure probability $\epsilon$ and a memoryless BSC with error probability $p$. While the erasure channel models the missing ratings, the BSC models noisy behavior of the buyers. The output of the channel, i.e.\ the observed rating matrix, is denoted by $\Y$ and its entries are in  $\{0, 1, e\}$, where $e$ denotes an erasure. We analyze the probability of error for a fixed rating matrix as well as the probability of error averaged over the rating matrices. We use the following probability law on the rating matrices. We assume that all row and column clusters have the same size $m_0$ and $n_0$ respectively, and the $rt$ constant blocks (of size $m_0 n_0$) contain i.i.d.\ Bernoulli 1/2 random variables. 

\section{Main Results}
\label{sect:main}

In Section \ref{sect:knownab}, we study the probability of error of the maximum likelihood decoder when the clusters ${\cal A}, {\cal B}$ are known. This result provides a lower bound on the cluster size that ensures diminishing probability of error. In Section \ref{sect:cluster}, we analyze the probability of error in identifying the clusters for a specific algorithm. These results  are integrated in Section \ref{sect:discuss} to obtain conditions on the cluster sizes for the overall probability of error to diminish to zero.  

\subsection{Probability of Error When Clustering is Known}
\label{sect:knownab}
In this section, we study the probability of error of the maximum likelihood decoder for a given rating matrix $\X$ when ${\cal A}$ and ${\cal B}$ are known. We denote this probability by $\peab(\X)$. We note that the ML decoder ignores the erasures, counts the number of 0's and 1's in each cluster $A_i \times B_j$, and takes a majority decision. Ties are resolved by tossing a fair coin. The following theorem provides simple upper and lower bounds on $\peab$.

\begin{theorem}
\label{thm:lb-ub1}
Let $0\leq p\leq 1/2$, and let 
\begin{align*}
p_1 & = \epsilon + 2  (1-\epsilon) \sqrt{p(1-p)}\\
G(u) & =1- \prod_{i=1,j=1}^{r,t} \left(1-u^{m_i n_j}\right).
\end{align*}
Then the probability of error of the ML decoder satisfies the following bounds:
\begin{equation}
\label{eq:lb-ub1}
G(\epsilon) \leq \peab(\X) \leq G(p_1).
\end{equation}
\end{theorem}

\proof We note that when $p=0$, we make an error in a cluster iff all the entries in the cluster are erased. Since the erasures in different clusters are independent, it follows that $\peab(\X) = G(\epsilon)$ for $p=0$. This gives the lower bound on $\peab(\X)$ for $p \geq 0$.  

Next we prove the upper bound.
Suppose in cluster  $A_i \times B_j$ we have $s$ non erased samples. Then the probability of correct decision in this cluster is given by
\begin{equation}
\label{eq:eijs}
\begin{split}
\prob(E_{i,j,s}^c) & = \sum_{q=0}^{\lfloor \frac{s}{2}\rfloor}\binom{s}{q} p^q(1-p)^{s-q}  \text{ if }s\text{ is odd}\\
& = \sum_{q=0}^{\frac{s}{2}-1}\binom{s}{q}p^q(1-p)^{s-q} \\
& \quad +
\dfrac{1}{2}\binom{s}{\frac{s}{2}}p^{\frac{s}{2}}(1-p)^{\frac{s}{2}} \text{ if }s\text{ is even}.
\end{split}
\end{equation}
Averaging over the number of non erased samples, the probability of correct decision in cluster $A_i \times B_j$ is given by
\begin{equation}
\label{eq:eij}
\prob(E_{i,j}^c) =  \sum_{s=0}^{m_i n_j}\binom{m_i n_j}{s}\epsilon^{m_i n_j-s}(1-\epsilon)^s \prob(E_{i,j,s}^c). 
\end{equation}
Since the erasure and BSC are memoryless
\begin{align}
\peab(\X) & = \prob\left(\cup_{i=1,j=1}^{r,t} E_{i,j} \right) \nonumber \\
& = 1 - \prod_{i=1,j=1}^{r,t} \prob\left(E_{i,j}^c\right). \label{eq:error}
\end{align}
Equations \eqref{eq:error}, \eqref{eq:eij}, and \eqref{eq:eijs} specify the probability of error. The desired upper bound is obtained by deriving an upper bound on $\prob(E_{i,j,s}^c)$. 
First we note that from \eqref{eq:eijs},
\[
1-\prob(E_{i,j,s}^c)  \leq \sum_{\lceil \frac{s}{2} \rceil}^s \binom{s}{q}p^q(1-p)^{s-q}. 
\]
But for $0\leq p \leq \frac{1}{2}$ and $q\geq \frac{s}{2}$, $p^q(1-p)^{s-q}\leq p^{\frac{s}{2}}(1-p)^{\frac{s}{2}}$. 
Substituting this in the previous equation, we have
\begin{equation}
	\prob(E_{i,j,s}^c) \geq 1-(2\sqrt{p(1-p)})^s. \label{eq:eijslb}
\end{equation}
From Equations \eqref{eq:eij} and \eqref{eq:eijslb}, we have $\prob(E_{i,j}^c) \geq 1- p_1^{m_in_j}$ and so from \eqref{eq:error}, $\peab(\X) \leq G(p_1)$. This completes the proof for the upper bound on $\peab(\X)$.
 \qed

Let us define the {\em smallest cluster size} as
\begin{eqnarray}
& s_{\ast}(\X) & := \min_{i,j} m_i n_j, \nonumber \\
\end{eqnarray}
and the {\em largest cluster size} as
\begin{eqnarray}
& s^{\ast}(\X) & := \max_{i,j} m_i n_j. \nonumber
\end{eqnarray}
The following corollary gives simpler bounds on $\peab(\X)$.
\begin{corollary}
\label{cor:lb-ub2}
Let $N_{\X}(s)$ be the number of clusters in $\X$ with exactly $s$ elements. Let
$$s_{\ast}(\X) \geq \frac{\ln(2)}{\ln(1/p_1)}.$$
Then
\begin{equation}
\label{eq:lb-ub2}
\begin{split}
\peab(\X) & \geq 1-\exp\left(-\sum_{s=1}^\infty N_{\X}(s) \epsilon^s\right), \\
\peab(\X) & \leq 1-\exp\left(-2\ln(2)\sum_{s=1}^\infty N_{\X}(s) p_1^s\right).
\end{split}
\end{equation}
In particular,
\begin{equation}
\label{eq:lb-ub3}
\begin{split}
\peab(\X) & \geq 1-\exp\left(- \frac{m n \epsilon^{s^{\ast}(\X)}}{s^{\ast}(\X)}\right), \\
\peab(\X) & \leq 1-\exp\left(- \frac{2\ln(2)m n p_1^{s_{\ast}(\X)}}{s_{\ast}(\X)}\right).
\end{split}
\end{equation}
\end{corollary}
\proof The proof is based on upper and lower bounds for $G(u)$. We note that $(1-x) \leq \exp(-x)$ and for $x \in [0,1/2]$, $1-x \geq \exp(-2\ln(2) x)$. Hence
\begin{align*}
\exp\left(-2\ln(2) \sum_{i=1,j=1}^{r,t} u^{m_i n_j} \right) & \leq \prod_{i=1,j=1}^{r,t} \left(1- u^{m_i n_j} \right) \\
& \leq \exp\left(-\sum_{i=1,j=1}^{r,t} u^{m_i n_j} \right).
\end{align*}
Where the first inequality holds for $u^{m_in_j}\leq \frac{1}{2}$. The sum in the exponent can be written in terms of the size of the clusters:
\[
\sum_{i=1,j=1}^{r,t} u^{m_i n_j} = \sum_{s=1}^\infty N_{\X}(s) u^s.  
\]
The bounds \eqref{eq:lb-ub2} now follow from Theorem \ref{thm:lb-ub1} by noting that $p_1^{m_i n_j} \leq 1/2$ for $s_{\ast}(\X) \geq {\ln(2)}/{\ln(1/p_1)}$.

To prove \eqref{eq:lb-ub3}, we note that 
\[
\sum_{s=1}^\infty N_{\X}(s) u^s \leq r t u^{s_{\ast}(\X)} \leq \frac{mn}{s_{\ast}(\X)} u^{s_{\ast}(\X)}.
\]
This gives the upper bound in \eqref{eq:lb-ub3}. The lower bound in \eqref{eq:lb-ub3} follows similarly.
\qed

We are interested in studying the cluster sizes that guarantee correct decisions asymptotically. Though \eqref{eq:lb-ub2} is tighter than \eqref{eq:lb-ub3}, the conditions arising out of \eqref{eq:lb-ub3} are cleaner and are stated below.
\begin{corollary}
\label{cor:cor2}
Suppose we are given a sequence of rating matrices of increasing size, that is, $mn \tends \infty$. Then the following are true.
\begin{enumerate}
\item If
\[
s_{\ast}(\X) \geq \frac{\ln(mn)}{\ln(1/p_1)} 
\]
then $\peab(\X) \tends 0$.
\item If 
\[
 s^{\ast}(\X) \leq \frac{(1-\delta)\ln(mn)}{\ln(1/\epsilon)}, \text{ for some }\quad \delta > 0,
\]
then $\peab(\X) \tends 1$.
\end{enumerate}
\end{corollary}
\proof First consider Part 1. From \eqref{eq:lb-ub3}, using $e^{-x} \geq 1-x$ we get
\[
\peab(\X) \leq \frac{2\ln (2) m n p_1^{s_{\ast}(\X)}}{s_{\ast}(\X)}.
\]
The RHS is a decreasing function of $s_{\ast}(\X)$ and hence substituting the lower bound on $s_{\ast}(\X)$ we get
\[
\peab(\X) \leq \frac{2\ln (2) \ln(1/p_1)}{\ln(mn)} \tends 0.
\]
For Part 2, we note that  $1-\exp\left(- m n \epsilon^{s^{\ast}(\X)} / s^{\ast}(\X)\right)$ is a decreasing function of $s^{\ast}(\X)$, and hence substituting the upper bound, we have from \eqref{eq:lb-ub3}
\[
\peab(\X) \geq 1-\exp\left(-\frac{\ln(1/\epsilon)(m n)^\delta}{(1-\delta)\ln(m n)}\right).
\]
But since $(m n)^\delta / \ln m n \tends \infty$, we have $\peab \tends 1$.

\subsection{Probability of Error in Clustering}
\label{sect:cluster}

Data mining researchers have developed several techniques for clustering data; see for example \cite[Chapter 4]{chakrabarti}. 
In this section, we analyze a simple polynomial time clustering algorithm. The algorithm clusters rows and columns separately. To cluster rows, we compute the normalized Hamming distance between two rows over commonly sampled entries. For rows $i,j$, this distance is:
\[
d_{ij} = \frac{1}{n} \sum_{k=1}^n 1\left(Y_{ik} \neq e, Y_{jk} \neq e\right) 1(Y_{ik} \neq Y_{jk}).
\]
If this is less than a threshold $d_0$, then the two rows are declared to be in the same cluster and otherwise they are declared to be in different clusters. We apply this process to all pairs of rows and all pairs of columns. Let $I_{ij}$ be equal to 1 if rows $i$, $j$ belong to the same cluster and let it be 0 otherwise. The algorithm gives an estimate:
\begin{equation*}
\hat{I}_{ij} = 
\begin{cases}
 & 1, \quad d_{ij} < d_0, \\
& 0, \quad d_{ij} \geq d_0.
\end{cases}
\end{equation*}
We are interested in the probability that we make an error in row clustering averaged over the probability law on the rating matrices described in Section \ref{sect:model}:
\[
\prcavg = \prob\left(\hat{I}_{ij} \neq I_{ij} \text{ for some } i,j\right).
\]
Once the rows are clustered, we can apply the same procedure to cluster columns.
Below we analyze the error probability $\prcavg$; the probability of error in finding column clusters has similar behavior.
\begin{theorem}
\label{thm:cluster}
Suppose we are given a sequence of rating matrices with \ $n \tends \infty$ and $t_n$ column clusters, such that $\limsup_{n \tends \infty} m/n < \infty $. Let 
\[
\mu := 2p(1-p)(1-\epsilon)^2, \quad \delta := (1-\epsilon)^2 (1-2p)^2 
\]
and choose $d_0 = \mu+\delta/3$. Then there exists a positive constant $C_0$ such that if $t_n > C_0 \ln(n)$, then
$\prcavg \tends 0$.
\end{theorem}
\proof We start by considering the choice of the threshold. When $i,j$ are in the same cluster,
\[
E[d_{ij} | I_{ij}=1, \X] = 2p(1-p)(1-\epsilon)^2 = \mu.
\]
When $i,j$ are in different clusters, let  $s_{ij}$ be the number of columns in which $i,j$ disagree. Then
\begin{equation*}
\begin{split}
& E[d_{ij} | I_{ij}=0, \X] \\
& = \frac{(1-\epsilon)^2}{n} \left[(p^2+(1-p)^2) s_{ij} + 2p(1-p) (n-s_{ij})\right]\\
& = \mu + \frac{s_{ij}}{n} \delta.
\end{split}
\end{equation*}
We choose 
\[
d_0 = \mu + \frac{\alpha_n}{n} \delta,
\]
where $\alpha_n$ is chosen below to obtain diminishing probability of error.

First we bound the probability of error when $I_{ij}=1$. We note that in this case $d_{ij}$ is the average of $n$ i.i.d.\ Bernoulli random variables with mean $\mu=2p(1-p)(1-\epsilon)^2$. Hence
\begin{align}
& \prob\left(\hat{I}_{ij} \neq 1 \big| I_{ij}=1, \X\right) \nonumber \\
& = \prob\left( d_{ij} - \mu \geq \frac{\alpha_n}{n} \delta \Big| I_{ij}=1, \X\right) \nonumber \\
& \leq \exp\left(- \frac{\delta^2 \alpha_n^2}{\mu n}\right) \label{eq:CB1}
\end{align}
where in the last step we have used the Chernoff bound \cite[Theorem 4.4, pp.\ 64]{mitzenmacher}.

Next consider the case $I_{ij}=0$. In this case, $d_{ij}$ is the average of $n-s_{ij}$ identically distributed Bernoulli random variables with mean $\mu$ and $s_{ij}$ identically distributed Bernoulli random variables with mean $\nu = (1-\epsilon)^2[p^2+(1-p)^2]$, all the random variables being independent.
So we have 
\begin{align}
& \prob\left(\hat{I}_{ij} \neq 0 \big| I_{ij}=0, \X\right) \nonumber \\
& \leq \frac{(1-\mu + \mu e^{\theta})^{n-s_{ij}}(1-\nu + \nu e^{\theta})^{s_{ij}}}{e^{nd_0\theta}}, \theta < 0 \label{eqn:ij1chern}\\
& \leq \exp \left(n(e^\theta -1)\beta_{ij}-nd_0\theta\right),\ \beta_{ij}=\mu + \delta \frac{s_{ij}}{n} \label{eqn:ij1lbnd}
\end{align}
where in \eqref{eqn:ij1chern} we have used the Chernoff bound and in \eqref{eqn:ij1lbnd} we have used the inequality $1+x \leq \exp(x)$. Choosing $\theta = \max(0, \ln(d_0/\beta_{ij}))$ (which is the optimal choice), we have
\begin{equation}
\label{eqn:ij1betterlbnd}
\begin{split}
& \prob\left(\hat{I}_{ij} \neq 0 \big| I_{ij}=0, \X\right) \nonumber \\
& \leq \begin{cases}  
\exp \left(n(d_0-\beta_{ij})+nd_0\ln\left(\frac{\beta_{ij}}{d_0}\right)\right) \text{ if }s_{ij}\geq \alpha_n \\
1 \text{ if } s_{ij}< \alpha_n.
\end{cases}
\end{split}
\end{equation}
\begin{equation}
	\label{eqn:ij1finallbnd}
\end{equation}
Note that for $s_{ij}\geq \alpha_n$, we have $0\leq (\beta_{ij}-d_0)/d_0\leq 1$, and so  
\begin{equation*}
\ln \left( \frac{\beta_{ij}}{d_0}\right) \leq \frac{\beta_{ij}-d_0}{d_0} - \frac{1}{6}\left(\frac{\beta_{ij}-d_0}{d_0}\right)^2. 
\end{equation*}
Substituting in \eqref{eqn:ij1betterlbnd}, if $s_{ij}\geq \alpha_n$, then

\begin{equation}
\label{eq:cond_bd}
\prob\left(\hat{I}_{ij} \neq 0 \big| I_{ij}=0, \X\right) \leq \exp\left(-\frac{\delta^2 (s_{ij}-\alpha_n)^2}{6(n\mu + \delta \alpha_n)}\right).
\end{equation}
Taking expectation in \eqref{eqn:ij1finallbnd} and using \eqref{eq:cond_bd}, we get,
\begin{align*}
& E\left[\prob\left(\hat{I}_{ij} \neq 0 \big| I_{ij}=0, \X\right) \right] \\
& \leq \prob\left(s_{ij} \leq \alpha_n \right) + E\left[\exp\left(-\frac{\delta^2 (s_{ij}-\alpha_n)^2}{6(n\mu + \delta \alpha_n)}\right)\right] \\
& \quad =: T_1 + T_2.
\end{align*}
We note that $s_{ij} = n_0 X$, where $X$ is Binomial($t_n$,1/2). Thus $E[s_{ij}]=n_0 t_n/2=n/2$ and $\text{var}\{s_{ij}\} = n n_0/4$. Thus if $n_0=o(n)$, then $s_{ij}$ concentrates around its mean. Hence to get a diminishing $T_1$, we choose
$\alpha_n = n/3$. Then
\begin{align}
T_1 & =P\left(s_{ij} \leq \frac{n}{3}\right) = P\left(X \leq \frac{t_n}{3}\right) \nonumber \\
& \leq P\left(|X-t_n/2| \geq \frac{t_n}{6}\right) \nonumber \\
& \leq  2 \exp \left(-\frac{t_n}{54}\right) \label{eq:CB2}
\end{align}
where we have used the Chernoff bound \cite[Corollary 4.6, pp.\ 67]{mitzenmacher}.

Substituting for $\alpha_n$ in $T_2$, we see that for a suitable positive constant $c$,
\begin{align}
T_2 & = E\left[\exp\left(-c n_0\frac{(X- t_n/3)^2}{t_n}\right)\right] \nonumber \\
& = \sum_{s=0}^{t_n} \binom{t_n}{s} 2^{-t_n} \exp\left(-c n_0\frac{(s-t_n/3)^2}{t_n}\right) \nonumber \\
& = \sum_{|s-t_n/3| > t_n/9}   \binom{t_n}{s} 2^{-t_n} \exp\left(-c \frac{n_0(s-t_n/3)^2}{t_n}\right) \nonumber \\
& \quad + \sum_{|s-t_n/3| \leq t_n/9}   \binom{t_n}{s} 2^{-t_n} \exp\left(-c \frac{n_0(s-t_n/3)^2}{t_n}\right) \nonumber\\
& \leq \exp\left(-c n/81\right) + \sum_{|s-t_n/3| \leq t_n/9}  2^{-t_n}  2^{t_n h(s/t_n)} \nonumber \\
& \leq \exp\left(-c n/81\right) + t_n 2^{-t_n(1-h(4/9))}. \label{eq:f-1}
\end{align}
From \eqref{eq:CB2} and \eqref{eq:f-1}, it follows that 
\begin{equation}
\label{eq:f0}
E\left[\prob\left(\hat{I}_{ij} \neq 0 \big| I_{ij}=0, \X\right) \right] \leq	T_1+T_2 \leq nc_1 \exp(-c_2 t_n). 
\end{equation}
where $c_1, c_2$ are positive constants. Since there are only $m(m-1)/2$ pairs of rows, the desired result follows. \qed

\noindent
{\bf Remark:} If we consider the probability of error in clustering for a fixed rating matrix, then to get diminishing probability of error asymptotically, we need
\[
m_0 n_0 > C \sqrt{mn\ln(m)\ln(n)}.
\]

\subsection{Estimation Under Unknown Clustering}
\label{sect:discuss}
In this section, we consider our full problem - estimation of the underlying rating matrix from noisy, sparse observations when clustering is not known. Our result is the following.

\begin{theorem}
\label{thm:final}
Consider the collection of block constant matrices with the probability law described in Section \ref{sect:model}. Let $m=\beta n$, $\beta >0$ fixed. Then there exist constants $C_i$, $1\leq i \leq 4$ such that the following holds for $t > C_3 \ln(n)$, $r > C_4 \ln(m)$.
\begin{enumerate}
\item If $m_0 n_0 \leq C_1 \ln(mn)$, then for any estimator of $\X$, $\pe \tends 1$ as $n \tends \infty$.
\item Consider an estimator which first clusters the rows and columns using the algorithm described in Section \ref{sect:cluster} and then uses ML decoding as in Section \ref{sect:knownab} assuming that the clustering is correct. If $m_0 n_0 \geq C_2 \ln(mn)$, then for this algorithm $\pe \tends 0$ as $n \tends \infty$.
\end{enumerate}
\end{theorem}
\proof
When ${\cal A}, {\cal B}$ are known, then under our model all feasible rating matrices are equally likely. Hence the ML decoder gives the minimum probability of error and so we have
$
\pe \geq E[\peab(\X)].
$
To prove Part 1), we next lower bound $E[\peab(\X)]$.
Let $T$ be the event that $s^{\ast}(\X)>m_0n_0$. We note that $\X \in T$ iff for some pair of row clusters all the $t$ column clusters have been generated equal or for some pair of columns all the $r$ row clusters have been generated equal. Using the union bound, we get that,
\begin{equation}
\prob (T) \leq \frac{\binom{r}{2}}{2^t} + \frac{\binom{t}{2}}{2^r} \leq m^2 2^{-t}+n^2 2^{-r}.
\end{equation}
We choose $C_1$, $C_2$ to ensure that the above bound decays to zero and hence $\prob (T) \tends 0$. Now,
\[
E[\peab(\X)]  \geq  E[\peab(\X); T^c]. 
\]
But on the event $T^c$, $s^{\ast}(\X)=m_0 n_0$ and from the lower bound in \eqref{eq:lb-ub3} we get
\begin{equation}
\begin{split}
&\pe \geq E[\peab(\X)] \geq \\
&(1-\prob (T))\left[1-\exp\left(- \frac{\ln(1/\epsilon)(m n)^\delta}{(1-\delta)\ln(m n)}\right)\right]
\end{split}
\end{equation}
which $\tends 1$ as $mn \tends \infty$. This proves Part 1).

Next we prove Part 2).
Let $D$ denote the event that the clustering is identified correctly. We note that the probability of error in estimating $\X$ averaged over the probability law on the block constant matrices satisfies 
\begin{equation*}
\begin{split}
\pe & \leq E\left[\peab(\X) \prob(D)+ \prob(D^c)\right] \\
& \leq E\left[\peab(\X)\right]  + \left(\prcavg + \bar{P}_{e,cc}\right)
\end{split}
\end{equation*}
where $\bar{P}_{e,cc}$ is the probability of error in column clustering. The desired result follows from Part 1) of Corollary \ref{cor:cor2}, and Theorem \ref{thm:cluster}. \qed

\newpage

\noindent
{\bf Remark:} The above result states that for a fixed $p, \epsilon$, the smallest cluster size that leads to zero error asymptotically is $O(\ln(mn))=O(\ln(n))$. When $p=0$, then we can also apply the method in \cite{recht2} to our model, and this yields a smallest cluster size of $O(n^{1/2}(\ln(n))^2)$, which is strictly worse than our result.  

\noindent
{\bf Remark:} In \cite{montanari}, the focus is on rating matrices of rank $O(1)$ and $\epsilon = c/n$, which leads to $O(n)$ observations. For our model, $O(1)$ rank corresponds to a cluster size of $\Theta(mn)$, and for $\epsilon = c/n$, our algorithm can be seen to give zero error asymptotically for any fixed rating matrix.

\section{Conclusion}
\label{sect:con}
We considered the problem of estimating a block constant rating matrix. 
The observed matrix is obtained through unknown relabeling of the rows and columns of the underlying matrix, followed by an error and erasure
channel. 
Our probability of error analysis showed that 
if
the number of row clusters and the number column clusters are
$\Omega(\ln (m))$ and $\Omega(\ln (n))$ respectively, then the matrix can be
clustered and estimated with vanishing probability of error if
the cluster sizes are $\Omega(\ln (mn))$. 

\section{Acknowledgments}
The work of Onkar Dabeer was supported by the XI Plan Project from
TIFR and the Homi Bhabha Fellowship. The work of Bikash Kumar Dey was supported by Bharti Centre
for Communication in IIT Bombay.

\end{document}